\documentclass[aps,pra,twocolumn,amsmath,floatfix,longbibliography,eprint=false]{revtex4-1}
\usepackage[utf8]{inputenc}
\usepackage{amsmath}
\usepackage{amsfonts}
\usepackage{amssymb}
\usepackage{graphicx}
\usepackage{hyperref}
\usepackage{color}

\newcommand{\cre}[1]{\hat{#1}^{\dagger}}

\begin{document}

\title{Phonon-mediated exciton capture in Mo-based transition metal dichalcogenides}

\author{F.~Lengers}
\email{Frank.Lengers@wwu.de}
\affiliation{Institut f\"ur Festk\"orpertheorie, Universit\"at M\"unster,
Wilhelm-Klemm-Str.~10, 48149 M\"unster, Germany}

\author{T.~Kuhn}
\affiliation{Institut f\"ur Festk\"orpertheorie, Universit\"at M\"unster,
Wilhelm-Klemm-Str.~10, 48149 M\"unster, Germany}

\author{D.~E.~Reiter}
\affiliation{Institut f\"ur Festk\"orpertheorie, Universit\"at M\"unster,
Wilhelm-Klemm-Str.~10, 48149 M\"unster, Germany}

\date{\today}

\begin{abstract}
Localized excitons play a vital role in the optical response of monolayers of transition metal dichalcogenides and can be exploited as single photon sources for quantum information technology. While the optical properties of such localized excitons are vastly studied, the ultrafast capture process of delocalized excitons into localized potentials is largely unexplored. We perform quantum kinetic calculations of exciton capture via acoustic and optical phonons showing that efficient capture takes place on an ultrafast time scale. The polaron formation in the low-temperature limit leads to higher-energy excitons which can then be efficiently trapped. We demonstrate that the interplay of acoustic and optical phonons leads to an efficient broadening of energy-selection rules. Our studies provide a deep understanding of the carrier trapping from two-dimensional materials into zero-dimensional potentials.
\end{abstract}


\maketitle
\section{Introduction}
Monolayers of transition metal dichalcogenides (TMDCs) have emerged as attractive candidates for novel optoelectronic applications due to their direct bandgap in the ultimate limit of a two-dimensional (2D) system \cite{Splendiani10,Mak10,Wang18,Mueller18}. Defects and other spatial energy fluctuations strongly affect optical and transport properties of excitons due to the large surface-to-volume ratio of monolayers \cite{Niehues18,Perea-Causin19,Zipfel20,Greben20,Chu20}. Localized excitons occur naturally, e.g., on edges \cite{Tonndorf15}, due to encapsulated air between monolayer and substrate \cite{Carmesin19,Darlington20} or atomic defects \cite{Greben20}. On the other hand local trapping potentials in  TMDCs have been deliberately created by means of strain or extrinsic atom radiation \cite{Tonndorf15,Fox15,Kern16,Branny17,Klein17,Tripathi18,Klein19}. Using the recombination of the carriers trapped in these potentials single photon emission can be achieved which may be used in quantum information processing \cite{Tonndorf15,Srivastava15,Klein20}. 

\begin{figure}[t!]
\includegraphics[width=\columnwidth]{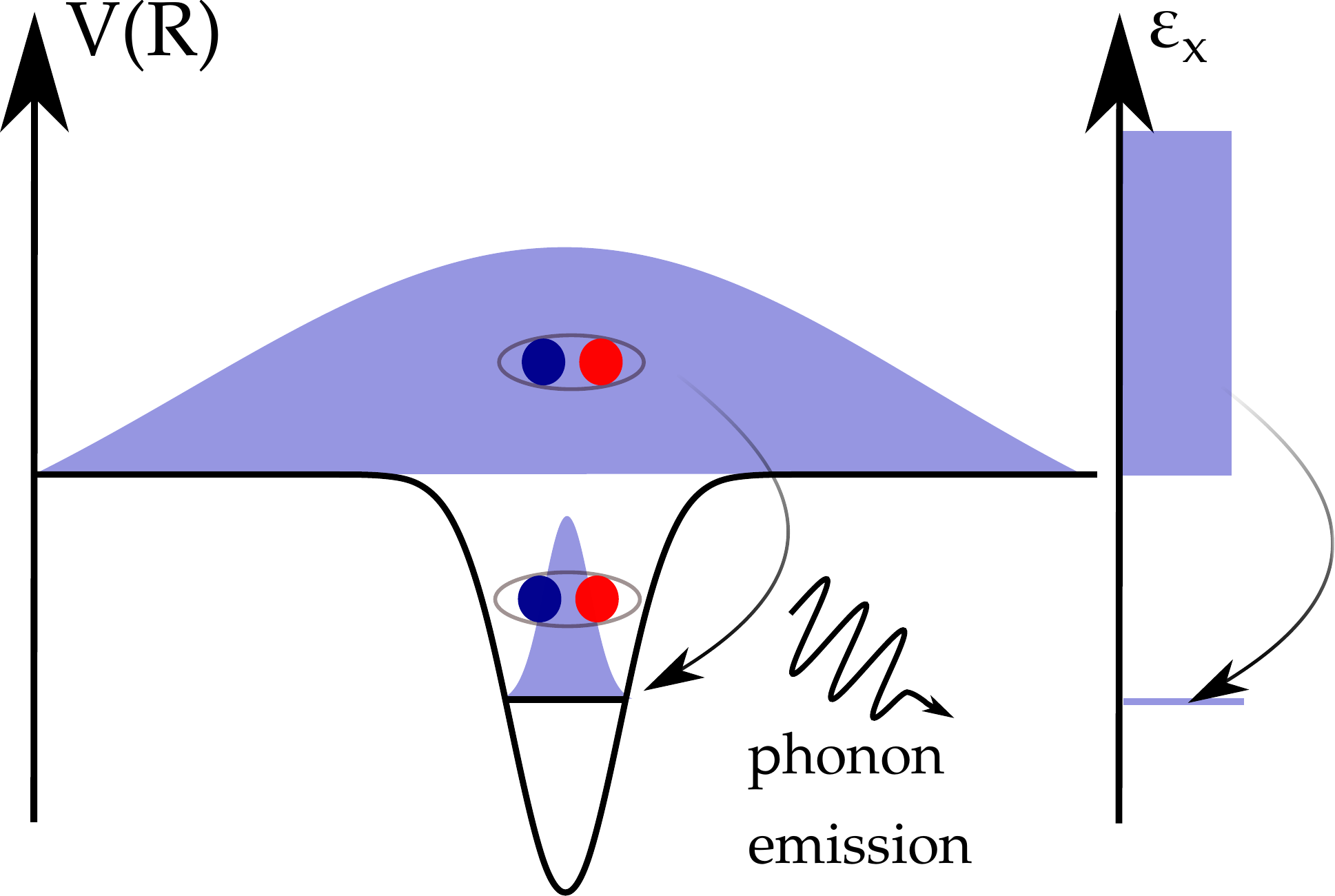}
\caption{Sketch of exciton capture as transitions from delocalized states within an energetic continuum to discrete localized states. The left part sketches the effective potential for the center-of-mass movement of the bound exciton (bound electron-hole pair sketched as blue and red dot). The right part illustrates the resulting energy spectrum.}
\label{Fig_introduction}
\end{figure}

In this paper, we theoretically investigate how the spatiotemporal dynamics of the carrier capture into a localized potential takes place. A special focus lies on the role of the phonons, which are mediating the carrier capture. We will further highlight the role of the incoherent excitons, which play a vital part in the capture. A sketch of the process is shown in Fig.~\ref{Fig_introduction} with the delocalized carriers in the 2D plane being captured into a localized potential. Clearly, the inhomogeneous nature and different dimensions involved in the process play an important role in describing the capture adequately. 

For simulations of spatiotemporal transport of excitons disorder is usually treated in an averaged way leading to a contribution to the diffusion constant due to scattering \cite{Kato16,Kulig18,Perea-Causin19}. Trapping of excitons has been treated by semiclassical rate equations \cite{Feierabend19}. It is known from semiconductor quantum dots in conventional semiconducting materials like GaAs that carrier capture is actually a complicated process which combines non-Markovian renormalizations broadening strict energy selection rules \cite{Thraenhardt00,Glanemann05,Muljarov07} and non-trivial spatiotemporal dynamics on a sub-picosecond time scale \cite{Reiter07,Rosati17,Rosati18}. In view of this, a spatially resolved quantum kinetic treatment of exciton capture is desirable for TMDCs, especially because of the strong exciton-phonon interaction in these systems usually leading to ultrafast exciton scattering \cite{Brem18,Lengers20,Chow17,Moody15,Ovesen19,Selig18,Zheng17}. 

Here, we derive quantum kinetic equations of motion in the density matrix formalism for exciton dynamics after optical excitation in the presence of a localization potential and exciton-phonon interaction. By formulating a non-diagonal density matrix within the exciton picture we are able to consider the spatial dynamics and can thereby study local density modifications in the vicinity of exciton traps. We will discuss the role of phonons in the capture process accounting for both optical and acoustic phonons.  Optical phonons are shown to lead to an efficient capture process, while acoustic phonons may stabilize the capture. We will discuss the effect of the polaron formation on the carrier capture and demonstrate that the resonance condition for capture via optical phonons is strongly broadened. 

\section{Theory}
\label{sec:theory}
In this section, we derive the equations of motion for the spatiotemporal dynamics of excitons in a quantum kinetic treatment. Due to the dimensionality of the problem, which consists of 2D delocalized carriers trapped into a 0D potential, a challenge lies in the computational feasibility. To slightly simplify the equations of motion we will separate the dynamics within delocalized states from the more interesting dynamics between delocalized and localized states. We treat the delocalized states in a time convolutionless approximation in the case of coherent excitons and in Born-Markov approximation in the case of incoherent excitons, while the capture dynamics between delocalized and localized states is treated in a full quantum kinetic calculation.

\subsection{Exciton states with a localized potential}
We will consider Mo-based TMDCs and treat the system in the exciton basis with the exciton creation (annihilation) operator $\hat{P}^{\dagger}_{\mathbf{K},v}$ ($\hat{P}_{\mathbf{K},v}$) of center-of-mass momentum $\mathbf{K}$ with dispersion relation $E_{\mathbf{K}}=E_{\mathrm{1s}}+\frac{\hbar^2 K^2}{2M}$ with the exciton mass $M$ and the bound state energy $E_{\mathrm{1s}}$ in the valley $v$. In Mo-based TMDCs one can reduce the attention to the optically bright K and K' valleys in the low-temperature limit when excited resonantly due to the energetic separations to dark valleys, in contrast to W-based TMDCs \cite{Malic18,Wang18,Selig20}. When choosing circularly polarized light as the source of excitation, we can just consider one of these valleys and drop the valley index $v$. In addition we will consider a localized confinement potential for the excitons, such that the excitonic Hamiltonian reads
\begin{align*}
	H_0=&\sum\limits_{\mathbf{K}}E_{\mathbf{K}}\cre{P}_{\mathbf{K}}\hat{P}_{\mathbf{K}}
	+\sum\limits_{\mathbf{K,Q}}\tilde{V}_{C,1s}(\mathbf{Q})\cre{P}_{\mathbf{K+Q}}\hat{P}_{\mathbf{K}}.
\end{align*}
In the exciton basis the confinement for the $1s$-excitons $\tilde{V}_{C,1s}$ is given by the Fourier transform of the effective exciton confinement ${V}_{C,1s}$. This is related to electron and hole confinement by
	\begin{align*}&{V}_{C,1s}(\mathbf{R})=\\
	&
	\int |\phi_{1s}(\mathbf{r})|^2\left[V_e\left(\mathbf{R}+\frac{m_h}{M}\mathbf{r}\right)+V_h\left(\mathbf{R}-\frac{m_e}{M}\mathbf{r}\right)\right]d^2r
	\end{align*}
	where $V_e$ and $V_h$ are the confinement potentials for electrons and holes, respectively, and $\phi_{1s}$ is the wave function of relative motion for the $1s$ excitons.\\
Considering not too strong confinement potentials and focussing on the energetically lowest states, we can restrict ourselves to the $1s$ excitons because then the admixture of other excitonic levels is unimportant due to the huge binding energy in TMDCs. Additionally, we take only  confinement potentials which vary slowly on the length scale of a unit cell, such that different valleys are not coupled by the confinement potential.

By solving the Schr\"odinger equation
\begin{align}
	\sum\limits_{\mathbf{Q}}\left(E_{\mathbf{K}}\delta_{\mathbf{Q},\mathbf{K}}+\tilde{V}_{C,1s}(\mathbf{K-Q})\right)\tilde{\Psi}_x(\mathbf{Q})=\epsilon_x\tilde{\Psi}_x(\mathbf{K}),
	\label{eq_schroedinger}
\end{align}
we can change into the energy eigenbasis, in which the excitonic Hamiltonian in the eigenbasis $x$ is given by
\begin{align}
	H_x=&\sum\limits_{x}\epsilon_x\cre{P}_x\hat{P}_x,
\end{align}
where the new exciton operators are $\cre{P}_x=\sum\limits_{\mathbf{K}}\tilde{\Psi}_x(\mathbf{K})\cre{P}_{\mathbf{K}}$.

\begin{figure}[t!]
\includegraphics[width=.65\columnwidth]{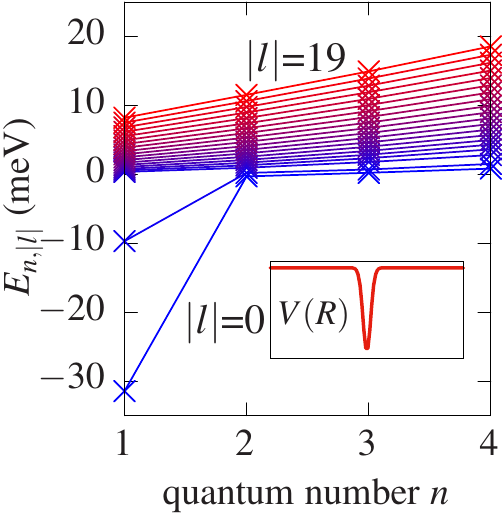}
\caption{Energy spectrum $E_{n,|l|}$ of a considered localization potential for angular momentum quantum numbers up to $|l|=19$ with the parameters $V_0=60~$meV and $\Delta_R=2~$nm. States of the same $|l|$ are connected by lines as guide to the eye.}
\label{Fig_potentials}
\end{figure}

Here we focus on strongly localized potentials, which are radially symmetric and of the form
\begin{equation} \label{eq_potential}
	{V}_{C,\mathrm{1s}}(\mathbf{R})=-V_0\exp\left(-\frac{R^2}{2\Delta_R^2}\right)
\end{equation}
with the potential depth $V_0$ and spatial width $\Delta_R$. We will assume $\Delta_R=2$~nm throughout the paper. An example of the potential with a depth of $V_0=60~$meV is shown in Fig.~\ref{Fig_potentials}, where the solutions of Eq.(\ref{eq_schroedinger}) are displayed. Because of the radial symmetry the wave functions can be labeled by the angular momentum quantum number $l$ such that ${\Psi}_{n,l}(\mathbf{R})={\Psi}_{n,l}(R)e^{il\phi_R}$. Thereby one finds doubly degenerate states for $|l|>0$. For such a small $\Delta_R$, we just find three bound states, one for $l=0$ and two degenerate states for $l=\pm 1$. For $E_x>0$ a quasi-continuum of delocalized states is found. Details of the numerical implications of radial symmetry and the evaluation of eigenstates can be found in App.~\ref{App_symmetry}.

\subsection{Interactions}
Following Refs.~\cite{Li13,Jin14}, for the phononic branches we only take into account one effective optical and one effective acoustic branch, where all optical phonons around the phononic $\Gamma$-point are combined with their mean optical frequency and the same is done for the three types of acoustic phonons. The optical phonon frequency is taken as $\hbar\omega_{\text{op}}=34.4\,$meV and the dispersion for the acoustic phonons is $\hbar\omega_{\text{ac}}=\hbar v_{s}Q$ with the sound velocity $v_s=4.1~\mathrm{{nm}/{ps}}$. The free phonon subsystem is described by
\begin{equation}
	H_{\mathrm{ph}}=\sum\limits_{i,\mathbf{Q}}\hbar\omega_{\mathbf{Q},i}\cre{b}_{\mathbf{Q},i}\hat{b}_{\mathbf{Q},i}
\end{equation}
 with the bosonic creation (annihilation) operator $\cre{b}_{\mathbf{Q},i}$ ($\hat{b}_{\mathbf{Q},i}$) of a phonon with momentum $\mathbf{Q}$ of branch $i=\{\text{op,ac}\}$. 
 
The Hamiltonian of the exciton-phonon interaction is 
\begin{equation}
	H_{x-\mathrm{ph}}=\sum\limits_{x,x',\mathbf{Q},i}g_{\mathbf{Q},i}K_{x\mathbf{Q}x'}\cre{P}_{x}\hat{P}_{x'}\left(\hat{b}_{\mathbf{Q},i}+\cre{b}_{-\mathbf{Q},i}\right)\,.
\end{equation}
The exciton-phonon interaction coupling matrix element $g_{\mathbf{Q},i}$ is derived from the difference between electron and hole coupling, including the form factor for the excitonic system. We here consider only the deformation potential interaction for both optical and acoustic phonons. The parameters for the deformation potential coupling are taken from Ref.~\cite{Lengers20}. By transforming into the energy eigenbasis the additional factor $K_{x,\mathbf{Q},x'}$ occurs with
\[
 	K_{x\mathbf{Q}x'}=\int {\Psi}_x^*(\mathbf{R}){\Psi}_{x'}(\mathbf{R})e^{i\mathbf{Q\cdot R}}d^2R.
\]

In our simulations, we will consider excitons excited by an optical laser pulse (typically a $\delta$-pulse) in dipole approximation. The interaction of the excitons with a classical light field is then determined by the dipole matrix element $\mathbf{M}_{1s}$ projected onto the electric field $\mathbf{E}(\mathbf{R},t)$. Assuming right-circularly polarized light, the exclusive excitation of the $K$ valley is a good approximation \cite{Wang18}. In rotating wave approximation the relevant part of the electric field is given by $\mathbf{E}(\mathbf{R},t)=\mathbf{e}_+ u(\mathbf{R})E(t)$ with $\mathbf{e}_+=1/\sqrt{2}(\mathbf{e}_x+i\mathbf{e}_y)$ and the spatial and temporal fields $u(\mathbf{R})$ and $E(t)$, respectively, \cite{Juergens20} such that the coupling strength is given by $M_{1s}=\mathbf{M}_{1s}\cdot\mathbf{e}_+$. In the energy eigenbasis the light-matter interaction then reads
\begin{equation}
	H_{x-l}=-\sum\limits_x M_{x}E(t)\cre{P}_x-\sum\limits_x\left(M_{x}E(t)\right)^*\hat{P}_x,
\end{equation}
where we have introduced the effective dipole matrix element
\[
	M_x=M_{\mathrm{1s}}\int {\Psi}_x^*(\mathbf{R})u(\mathbf{R}) d^2R\propto \sum\limits_{\mathbf{K}} \tilde{\Psi}_x^*(\mathbf{K})\tilde{u}(\mathbf{K}) \,.
\] 
where the spatial envelope $u(\mathbf{R})$ is assumed to be only varying on a mesoscopic scale such that it only acts on the center-of-mass envelope $\Psi_x$.
Note that in the last expression $\tilde{\Psi}_x(\mathbf{K})$ and $\tilde{u}(\mathbf{K})$ are the spatial Fourier transformations of their real-space counterparts such that only $\mathbf{K}=\mathbf{0}$ excitons are excited for a spatially homogeneous light field.

\subsection{Dynamic variables}
To study the spatiotemporal dynamics of the exciton capture, we will consider the time evolution of the density matrix given by 
\[
	\rho_{xx'}=\langle\cre{P}_x\hat{P}_{x'}\rangle.
\]
The full density matrix can be divided into a coherent and an incoherent part \cite{Thraenhardt00,Siantidis01,Selig18,Brem18}. The coherent occupation of excitons reads
\begin{equation} \label{eq_Ncoh}
	N_{\text{coh}} = \sum_{x}  \langle\cre{P}_x\rangle \langle \hat{P}_{x}\rangle  = \sum_{x} |{P}_x|^2 \, ,
\end{equation}
while the incoherent part of the density matrix is given by
\begin{equation} \label{eq_Nxx}
	N_{xx'}=\delta\langle\cre{P}_x\hat{P}_{x'}\rangle =\langle\cre{P}_x\hat{P}_{x'}\rangle-P_x^*P_{x'}
\end{equation}
resulting in the incoherent exciton occupation
\begin{equation} \label{eq_Nincoh}
	N_{\text{incoh}} = \sum_{x}N_{xx} \,.
\end{equation}
The total exciton occupation is then $N=N_{\text{coh}}+N_{\text{incoh}}$. Because we here consider only the low-density case, we neglect all higher correlations of exciton operators \cite{Katsch19}.

\subsection{Equations of motion for the polarization}
We start by setting up the equations of motion for the exciton polarization $P_x$, which is already sufficient to calculate the coherent exciton occupation $N_{\text{coh}}$. Using the Heisenberg equation of motion we obtain in second Born approximation (2BA)
\begin{align}
	i\hbar\frac{d}{dt}P_{x}&= \epsilon_{x}P_{x}-M_{x}E(t) \label{eom_coh_pol} \\
		&+\sum\limits_{x',\mathbf{Q},i}K_{x\mathbf{Q}x'}
			\left[S^{(P,+)}_{x',i}(\mathbf{Q})+S^{(P,-)}_{x',i}(\mathbf{Q})\right]\nonumber \\
	i\hbar\frac{d}{dt}S_{x',i}^{(P,+)}(\mathbf{Q})&=
		\left(\epsilon_{x'}-\hbar\omega_{\mathbf{Q},i}\right)S_{x',i}^{(P,+)}(\mathbf{Q})
		\label{eom_S_p}\\
	&+n_{\mathbf{Q},i}\sum\limits_{x}|g_{\mathbf{Q},i}|^2\left(K_{x\mathbf{Q}x'}\right)^* P_{x}\nonumber\\
	i\hbar\frac{d}{dt}S_{x',i}^{(P,-)}(\mathbf{Q})&=
	\left(\epsilon_{x'}+\hbar\omega_{\mathbf{Q},i}\right)S_{x',i}^{(P,-)}(\mathbf{Q})\\
	&+(1+n_{\mathbf{Q},i})\sum\limits_{x}|g_{\mathbf{Q},i}|^2\left(K_{x\mathbf{Q}x'}\right)^* P_{x},\nonumber
\end{align}
where we have introduced the phonon-assisted polarizations 
\begin{eqnarray*}
	S_{x',i}^{(P,+)}(\mathbf{Q}):=g_{\mathbf{Q},i}\langle \hat{P}_{x'}\hat{b}_{-\mathbf{Q},i}^{\dagger}\rangle; \,\,
	S_{x',i}^{(P,-)}(\mathbf{Q}):=g_{\mathbf{Q},i}\langle \hat{P}_{x'}\hat{b}_{\mathbf{Q},i}\rangle 
\end{eqnarray*}
and the thermal Bose distributions of phonons $n_{\mathbf{Q},i}=\langle \cre{b}_{\mathbf{Q},i}\hat{b}_{\mathbf{Q},i}\rangle$. 

In 2BA all 2-phonon-assisted correlations have been omitted \cite{Rossi02}. However, this level of accuracy is most likely insufficient to correctly describe the dynamics. For optical signals and dephasing dynamics, we have shown in Ref.~\cite{Lengers20} that 2BA produces unphysical results, especially for high temperatures, due to the strong exciton-phonon interaction in TMDCs. A possible solution is to take into account higher order correlations, but due to the considered inhomogeneity, such an approach is numerically not feasible. 

To account for the influence of the higher order correlation and still have a numerically feasible approach, we split the states into localized and delocalized states. Using the localized potential from Eq.~\eqref{eq_potential}, all states with $\epsilon_x\ge 0$~meV can be assumed to be delocalized. For the delocalized states we then perform a time convolutionless (TCL) approach for the phonon-assisted polarizations. In Ref.~\cite{Lengers20}, such an approach was introduced and successfully used to describe polarization dynamics of TMDCs and the related phonon-assisted spectra in the homogeneous case. If $\epsilon_{x'}<0$, the states are considered as localized and we will here use quantum kinetic equations within a 2BA.

For the optical excitation, we set an initial condition for the polarization by assuming a $\delta$-like optical excitation in time approximating an optical pulse which is spectrally broader than the considered range of optically active states. We choose $t_0=0$ as the time of optical excitation. 

With this we obtain explicit formulas for the phonon-assisted polarizations for the delocalized states $\epsilon_{x'}\ge 0$ ($\mathbf{DL}$)

\begin{align}
	S_{x',i,\textbf{DL}}^{(P,+)}(t)
	&=
		n_{\mathbf{Q},i}\sum\limits_x |g_{\mathbf{Q},i}|^2\left(K_{x\mathbf{Q}x'}\right)^*P_x(t) \notag \\
		&\times\frac{e^{-\frac{i}{\hbar}(\epsilon_{x'}-\epsilon_x-\hbar\omega_{\mathbf{Q},i})t}-1}{-\frac{i}{\hbar}(\epsilon_{x'}-\epsilon_x-\hbar\omega_{\mathbf{Q},i})} 
	\label{eom_coh_pol_DL}
\end{align}
and for the localized states $\epsilon_{x'} < 0$ ($\mathbf{L}$)
\begin{align}
	S_{x',i,\textbf{L}}^{(P,+)}(t)
	&=
		n_{\mathbf{Q},i}\sum\limits_x |g_{\mathbf{Q},i}|^2\left(K_{x\mathbf{Q}x'}\right)^* \notag \\
			&\times\int_0^t e^{-\frac{i}{\hbar}(\epsilon_{x'}-i\hbar\Gamma_{x'}-\hbar\omega_{\mathbf{Q},i})\tau}P_x(t-\tau)d\tau
	\label{eom_coh_pol_L}
\end{align}
and analogously for $S^{(P,-)}$. 

This special distinction between localized and delocalized states can be reasoned as follows: In the equation of motion for $P_x$ in Eq.~\eqref{eom_coh_pol} there is a sum over $x'$ and therefore a sum over the oscillating exponential in Eq.~\eqref{eom_coh_pol_DL}. In case of delocalized states, the sum over $x'$ is a sum over a continuum of energies which results in a finite memory depth and therefore converges to a simple dephasing rate at $t\rightarrow\infty$ (compare Ref.~\cite{Lengers20}). In the case of a discrete spectrum, the memory depth may be infinite in the case of optical phonons and a TCL approximation is not applicable anymore, which is linked to an infinite bath-correlation time \cite{Breuer02BOOK}. For a single excitonic level the TCL is exact \cite{Krummheuer02,Richter10}, but the coupling between multiple discrete levels by optical phonons may lead to instabilities in the TCL formulation. Here, the 2BA leads to stable dynamics.

The equation of motion in 2BA is additionally renormalized/damped by a damping rate $\Gamma_{x'}$, which approximately takes into account higher-order correlations \cite{Schilp94,Schilp95,Christiansen17,Lengers20}. The damping rate is defined as
\[
	\Gamma_x= \frac{1}{T_{\mathrm{av}}}\int_0^{T_{\mathrm{av}}}\gamma_x(t) dt
\]
		with 
\begin{align}
	\gamma_x(t) &=\frac{1}{\hbar^2}\sum\limits_{x',\mathbf{Q},i}|g_{\mathbf{Q},i}|^2|K_{x\mathbf{Q}x'}|^2 \notag \\
	& \times  \left[(1+n_{\mathbf{Q},i})\int_0^t \exp\left(-\frac{i}{\hbar}(\epsilon_{x'}-\epsilon_x+\hbar\omega_{\mathbf{Q},i})\tau\right)d\tau \right. \notag \\
	& +\left.n_{\mathbf{Q},i}\int_0^t \exp\left(-\frac{i}{\hbar}(\epsilon_{x'}-\epsilon_x-\hbar\omega_{\mathbf{Q},i})\tau\right)d\tau\right]
\end{align}
as derived in App.~\ref{App_gamma}. The definition of $\Gamma_x$ can be understood as an average contribution from higher-order correlations over a defined time period $T_{\mathrm{av}}$ (chosen as $2~$ps throughout the paper which matches the simulation duration). The averaging over a certain time is further motivated by two aspects. Most importantly, the rates $\gamma_x$ of delocalized states rapidly converge to a quasistationary value on a timescale of about $100~$fs \cite{Lengers20} such that an averaged value is a good approximation. The rate only differs from that quasistationary value on a timescale of tens to hundreds of ps being much larger than the timescale of any scattering event considered here. Secondly, the dominant effect of spectral broadening is linked to delocalized states as shown for quantum dots coupled to a wetting layer by optical phonons \cite{Seebeck05,Muljarov07} while the purely discrete contributions by localized states are approximately averaged out and therefore do not induce problems like in the TCL approximation as discussed above. We note that while one could in principle introduce the completed collision limit $T_{\mathrm{av}}\rightarrow\infty$ in the compuation of $\Gamma_x$ as done in Refs.~\cite{Schilp94,Schilp95}, we nevertheless do not perform this approximation since we showed in Ref.~\cite{Lengers20} that such an approximation for acoustic phonons may overestimate their effect on ultrashort timescales. We  note that a change of $T_{\mathrm{av}}$ in the range of picoseconds did not lead to a notable change of the numerical results. 
\subsection{Equations of motion for incoherent excitons}
Next, we consider the dynamics of the incoherent excitons. Since the dynamics of the occupation of coherent excitons is already given by the polarization, we here consider the dynamics of incoherent excitons $N_{xx'}$ [cf. Eq.~\eqref{eq_Nxx}]. Such incoherent excitons do not build up due to the optical excitation, which initializes only a coherent exciton density. Instead, exciton-phonon interaction results in scattering and, thus, destroys the coherence of the exciton, resulting in an incoherent exciton occupation. The corresponding equation of motion reads
	\begin{align*}
	\frac{d}{dt}N_{xx'}=-\frac{i}{\hbar}(\epsilon_{x'}-\epsilon_{x})N_{xx'}+\frac{d}{dt}N_{xx'}\Big|_{\mathrm{coh}}+\frac{d}{dt}N_{xx'}\Big|_{\mathrm{incoh}}.
	\end{align*}
The first scattering part $\frac{d}{dt}N_{xx'}\Big|_{\mathrm{coh}}$ describes the aforementioned scattering from coherent into incoherent excitons with
	\begin{align*}
	i\hbar&\frac{d}{dt}N_{xx'}\Big|_{\mathrm{coh}}=\\
	&\sum\limits_{\bar{x},\mathbf{Q},i}\left(K_{\bar{x}\mathbf{Q}x'}\right)^* P_{\bar{x}}\left(S_{x,i}^{(P,-)}(\mathbf{Q})+S_{x,i}^{(P,+)}(\mathbf{Q})\right)^*
\nonumber\\
-&\sum\limits_{\bar{x},\mathbf{Q},i}K_{\bar{x}\mathbf{Q}x} \left(P_{\bar{x}}\right)^*\left(S_{x',i}^{(P,-)}(\mathbf{Q})+S_{x',i}^{(P,+)}(\mathbf{Q})\right)
\nonumber.
	\end{align*}
It describes the temperature-dependent dephasing of the polarization which is known as the polarization to population transfer \cite{Kira06,Selig16} and also the build-up of the polaron. Note that the second effect is only described by a quantum kinetic formulation.

The scattering between incoherent excitons is summarized as
\begin{align*}
i\hbar
\frac{d}{dt}N_{xx'}\Big|_{\mathrm{incoh.}}=&\sum\limits_{\bar{x},\mathbf{Q},i} K_{x'\mathbf{Q}\bar{x}} \left(S_{x\bar{x},i}^{(N,-)}(\mathbf{Q})+S_{\bar{x}x,i}^{(N,-)*}(\mathbf{-Q})\right)
\nonumber\\
-&\sum\limits_{\bar{x},\mathbf{Q},i} K_{\bar{x}\mathbf{Q}x} \left(S_{\bar{x}x',i}^{(N,-)}(\mathbf{Q})+S_{x'\bar{x},i}^{(N,-)*}(\mathbf{-Q})\right).
\nonumber
\end{align*}
Here, we have defined the phonon-assisted exciton density matrix via
\begin{align*}
S_{x x',i}^{(N,-)}(\mathbf{Q})
=g_{\mathbf{Q},i}\delta\langle\hat{P}_x^{\dagger}\hat{P}_{x'}\hat{b}_{\mathbf{Q},i}\rangle
\end{align*}
with the correlation of incoherent excitons and phonons
\begin{align*}
	\delta\langle\hat{P}_x^{\dagger}\hat{P}_{x'}\hat{b}_{\mathbf{Q},i}\rangle
	&=\langle\hat{P}_x^{\dagger}\hat{P}_{x'}\hat{b}_{\mathbf{Q},i}\rangle\\
 &-\langle\hat{P}_{x}^{\dagger}\rangle\langle\hat{P}_{x'}\hat{b}_{\mathbf{Q},i}\rangle-\langle\hat{P}_{x'}\rangle\langle\hat{P}_{x}^{\dagger}\hat{b}_{\mathbf{Q},i}\rangle.
\end{align*}
The equation of motion of $S^{(N)}$ reads within 2BA
\begin{align}
  i\hbar\frac{d}{dt}S_{x x',i}^{(N,-)}(\mathbf{Q}) &=
\left(\epsilon_{x'}-\epsilon_{x}+\hbar\omega_{\mathbf{Q},i}\right)S_{x x',i}^{(N,-)}(\mathbf{Q})
\label{EOM_incoh_2BA}
 \\
 &+(1+n_{\mathbf{Q},i})|g_{\mathbf{Q},i}|^2\sum\limits_{\bar{x}}\left(K_{\bar{x}\mathbf{Q}x'}\right)^*N_{x\bar{x}}\nonumber\\
 &-n_{\mathbf{Q},i}|g_{\mathbf{Q},i}|^2\sum\limits_{\bar{x}}\left(K_{x\mathbf{Q}\bar{x}}\right)^*N_{\bar{x}x'}.
 \nonumber
\end{align}
Like before, we treat the dynamics within the delocalized states separately. This corresponds to the case if all states $x,x',\bar{x}$ in Eq.~(\ref{EOM_incoh_2BA}) are delocalized. Because all equations of motion are linear, the separation can be easily performed by
	$$S_{x x',i,\mathbf{DL}}^{(N,-)}(\mathbf{Q})=S_{x x',i,\mathbf{DL}}^{(N,-)}(\mathbf{Q})\Big|_{\mathbf{DL}}+S_{x x',i,\mathbf{DL}}^{(N,-)}(\mathbf{Q})\Big|_{\mathbf{L}}$$
	for $x,x'\in \mathbf{DL}$, $\mathbf{DL}$ being the subspace of delocalized states ($\epsilon_x\ge 0$). The variables $S_{\mathbf{DL}}^{(N,-)}(\mathbf{Q})\Big|_{\mathbf{X}}$ obey the equation of motion in Eq.~\eqref{EOM_incoh_2BA} where the summations on the right hand side are restricted to $\bar{x}\in\mathbf{DL}$ for $\mathbf{X}=\mathbf{DL}$ and to $\bar{x}\notin\mathbf{DL}$ for $\mathbf{X}=\mathbf{L}$. Thereby $S_{\mathbf{DL}}^{(N,-)}(\mathbf{Q})\Big|_{\mathbf{DL}}$ treats the scattering within the delocalized states, while $S_{\mathbf{DL}}^{(N,-)}(\mathbf{Q})\Big|_{\mathbf{L}}$ connects delocalized and localized states.

We treat the dynamics within the delocalized states within a Boltzmann approximation as done in several works on TMDCs in the homogeneous \cite{Brem18,Selig18,Ovesen19} and inhomogeneous case \cite{Perea-Causin19,Rosati20}.
Using a Born-Markov approximation for $S_{\mathbf{DL}}^{(N,-)}(\mathbf{Q})\Big|_{\mathbf{DL}}$ results in Boltzmann-like transition rates (omitting the Cauchy Principal value resulting in energy renormalizations) with
\begin{align*}
	&S_{x,x',i,\mathbf{DL}}^{(N,-)}(\mathbf{Q})\Big|_{\mathbf{DL}}=-i\pi(1+n_{\mathbf{Q},i})|g_{\mathbf{Q},i}|^2\\
	&\times\sum\limits_{\bar{x}\in\mathbf{DL}}\left(K_{\bar{x}\mathbf{Q}x'}\right)^*N_{x\bar{x}}\delta\left(\epsilon_{x'}-\epsilon_{\bar{x}}+\hbar\omega_{\mathbf{Q},i}\right)\\
	&+i\pi n_{\mathbf{Q},i}|g_{\mathbf{Q},i}|^2\sum\limits_{\bar{x}\in\mathbf{DL}}\left(K_{x\mathbf{Q}\bar{x}}\right)^*N_{\bar{x}x'}\delta\left(\epsilon_{\bar{x}}-\epsilon_{x}+\hbar\omega_{\mathbf{Q},i}\right).
	\end{align*}
All remaining correlations, i.e., $S_{x,x',i}^{(N,-)}$ for $x\notin\mathbf{DL}\vee x'\notin\mathbf{DL}$ and $S_{x,x',i,\mathbf{DL}}^{(N,-)}(\mathbf{Q})\Big|_{\mathbf{L}}$, are treated dynamically and can be abbreviated as
\begin{align}
 i\hbar\frac{d}{dt}S_{x x',i}^{(N,-)}(\mathbf{Q})=&
\left(\epsilon_{x'}-\epsilon_{x}+\hbar\omega_{\mathbf{Q},i}\right)S_{x x',i}^{(N,-)}(\mathbf{Q})
 \nonumber
 \\
 +&(1+n_{\mathbf{Q},i})|g_{\mathbf{Q},i}|^2\sum\limits_{\bar{x}\in {\mathbf{M}}_{x,x'}}\left(K_{\bar{x}\mathbf{Q}x'}\right)^*N_{x\bar{x}}
 \nonumber\\
 -&n_{\mathbf{Q},i}|g_{\mathbf{Q},i}|^2\sum\limits_{\bar{x}\in {\mathbf{M}}_{x,x'}}\left(K_{x\mathbf{Q}\bar{x}}\right)^*N_{\bar{x}x'}\nonumber\\
 -&i\hbar\left(\Gamma_{x'}+\Gamma_{x}^*\right)S_{x x',i}^{(N,-)}(\mathbf{Q}),
 \nonumber
 \end{align}
where 
	\begin{align*}
	{\mathbf{M}}_{x,x'}=
		\begin{cases}
		\left\{\bar{x}|\bar{x}\notin\mathbf{DL}\right\} & \text{if  } x\in\mathbf{DL} \wedge x'\in\mathbf{DL}\\
		\left\{\bar{x}\right\} & \text{else}
		\end{cases}
	\end{align*}
accounts for the fact that we treated transitions between delocalized states by a Markovian approximation. In these equations of motion we added a damping/renormalization $\Gamma_x$ in analogy to Eq.~(\ref{eom_coh_pol_L}) to approximately account for higher-order correlations. These dampings mainly correct negativities in the density matrix which can occur in 2BA. We never experienced visible negativities due to the Born-Markov approximation within the delocalized states even though in the case of a non-diagonal density matrix the positivity is not strictly guaranteed \cite{Taj09,Pepe12}. On the other hand the introduced $\Gamma_x$ lead to the violation of energy conservation when considering the closed system of excitons and phonons \cite{Schilp95}. We are nevertheless not interested in the phonon dynamics and apart from that treat scattering within the delocalized states in a Born-Markov approximation because these states could potentially lead to a large increase in total energy. For the localized states we include $\Gamma_x$ because some clearly visible negativities occur in 2BA which are then weakened by inclusion of $\Gamma_x$ (see App.~\ref{App_negativities}). Additionally we already showed that a damping of correlations proved to be advantageous in the case of homogeneous dephasing dynamics as studied in Ref.~\cite{Lengers20}.

 All in all we have presented a theory describing the exciton-phonon interaction after ultrafast optical excitation to study the spatiotemporal exciton dynamics in an inhomogeneous system. In order to focus on capture processes from delocalized to localized states, we separated the dynamics within the delocalized subspace from the full state space. 

 It is interesting to briefly compare the presented theory of carrier capture to theories applied to conventional III--V semiconductor heterostructures with an embedded quantum dot potential \cite{Seebeck05,Glanemann05}. The presented theory is similar in the sense that carrier-phonon correlations are treated dynamically to describe non-Markovian features of the capture process. Specific to the present case is, that the carrier dynamics is treated within the excitonic picture, therefore all Coulomb correlations are treated exactly in the low-density limit. This is crucial for the case of TMDCs because of the huge exciton binding energy. We here focus on localized excitons which we assume to be formed by confinement of strongly bound bulk $1s$ excitons such that the capture into the energetically lowest excitons predominantly occurs from the free excitonic $1s$ state. Direct capture from uncorrelated electron-hole pairs is unlikely because of the large exciton binding energy. In the case of III--V semiconductors, the exciton binding energy is much smaller. Accordingly, already localized single particle energies (without considering Coulomb interaction) lie below or in the range of the bulk exciton binding energy, such that electron and hole capture may be treated separately. This has consequences for the capture process, because uncorrelated electron-hole pairs couple strongly by polar phonon coupling, while polar coupling is ineffective for strongly bound excitons due to large compensation of the charge density. This means that the usually dominant Fröhlich coupling of longitudinal optical phonons is weak for the localized states of $1s$ excitons considered here.

\section{Dynamics after optical excitation}
\label{sec:results_coh}
Having set-up the equations of motion, we now study the phonon mediated carrier capture from the 2D delocalized states into the 0D potential [cf. Eq.~\eqref{eq_potential}]. As material we consider $\mathrm{MoSe_2}$ with all parameters as in Ref.~\cite{Lengers20} and focus on the low-temperature limit by setting $T=0~$K throughout the paper. To simplify the numerical calculations, we restrict ourselves here to radially symmetric problems.  We start by looking at resonant excitation of the $1s$ exciton as used for experiments focussing, e.g., on valleytronic applications \cite{Kioseoglou12,Zhu14,Lundt19}.

To initialize coherent excitons in the system, we model an excitation, which is quasi resonant to the $1s$ exciton with an ultrafast pulse. The temporal shape of the pulse is taken as a $\delta$-function. The spatial form of the pulse is taken to be Gaussian with
\[
 u(\mathbf{R})=\exp\left(-R^2/\Delta_E^2\right)
\] 
with a spatial width of $\Delta_E=30~$nm. We thereby initialize coherent excitons proportional to $M_x\propto \int \Psi_x^*(\mathbf{R})u(\mathbf{R})d^2R$ which mainly results in excitation of delocalized excitons with energies around $\epsilon_x=0~$meV, because those have broad wavefunctions and only slow spatial oscillations.
For $\Delta_R=2~$nm  the fraction of directly excited localized excitons is low, since the pulse is much broader than the localization potential. Therefore, we initialize a situation where mainly delocalized excitons are excited and the capture from delocalized to localized states can be studied well.

\subsection{Capture into localized potential}
\label{Sec_coh_dephase}

\begin{figure}[t!]
\includegraphics[width=\columnwidth]{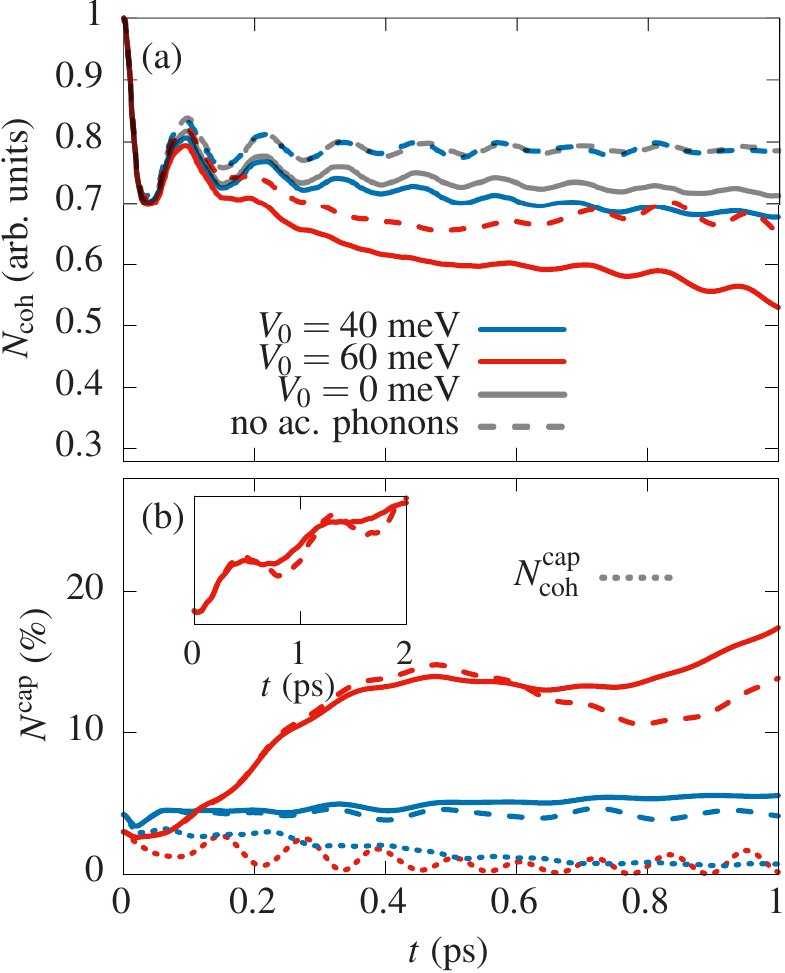}
\caption{(a) Dynamics of the occupation of coherent exciton $N_{\mathrm{coh}}$ for no potential ($V_0=0$~meV, gray lines, ) and with localized potential with depth $V_0=40$~meV (blue lines) and $V_0=60$~meV (red lines). Dashed lines are calculations without acoustic phonons.  (b) Fraction of localized excitons $N^{\text{cap}}$ for the two potentials with $\left.\frac{d}{dt}N_{xx'}\right|_{\mathrm{incoh}}=0$. Dotted lines show the fraction of coherent, captured excitons $N^{\text{cap}}_{\text{coh}}$. The inset shows the dynamics for $V_0=60~$meV up to $2$~ps.}
\label{Fig_coh_dephase}
\end{figure}

Figure~\ref{Fig_coh_dephase}(a) shows the dynamics of the coherent exciton occupation $N_{\mathrm{coh}}$ [Eq.~\eqref{eq_Ncoh}]. Let us first focus on the case without potential, i.e., $V_0=0$~meV (gray lines). After the initialization an initial drop of the occupation occurs followed by an oscillation of $N_{\mathrm{coh}}$. This behavior can be linked to the formation of the polaron mainly due to interaction with optical phonons. The formation of the polaron is accompanied by a loss of coherence, i.e., by formation of incoherent excitons. Figure~\ref{Fig_coh_dephase} shows that about $25\%$ of the optically excited (coherent) excitons are converted to incoherent excitons on the timescale of about $100~$fs, taking into account that the total number of excitons is conserved under exciton-phonon interaction with $\frac{d}{dt}(N_{\text{coh}}+N_{\text{incoh}})=0$. The dashed line shows the calculation with only optical phonons (no acoustic phonons). The scattering with acoustic phonons leads to an additional loss of coherent excitons on a slower time scale compared to optical phonons, transforming even more coherent excitons into incoherent ones. This is related to the fact that the exciton-phonon correlations of acoustic phonons relax to the Markovian limit on a much longer time scale than for optical phonons \cite{Lengers20}.

Now we want to study the phonon-induced capture of excitons into the localized potential. Therefore, we plot the fraction of captured carriers, defined by
\[
	N^{\mathrm{cap}}=\sum\limits_{x\in \mathbf{L}} \langle\cre{P}_x\hat{P}_{x} \rangle /N = N^{\mathrm{cap}}_{\text{coh}} + N^{\mathrm{cap}}_{\text{incoh}}\, ,
\] 
in Fig.~\ref{Fig_coh_dephase}(b). Note that both coherent and incoherent excitons are included in the fraction of captured excitons. To exclude very weakly localized states (being essentially delocalized over the simulation range), we restricted the summation of localized excitons to states with $\epsilon_x<-1~$meV. In this calculation, we have switched off $\left.\frac{d}{dt}N_{xx'}\right|_{\mathrm{incoh}}$ to focus on the capture from coherent excitons. We will include scattering between incoherent excitons in the next section. 

In the first example we take a potential depth of $V_0=40~$meV, which has the lowest bound state at $\epsilon_1=-20~$meV.  The dynamics for this potential (blue curves in Fig.~\ref{Fig_coh_dephase}) is very similar to the case without potential. When we look at Fig.~\ref{Fig_coh_dephase}(b), indeed, we find that almost no capture into the potential takes place. This is because of the lack of scattering due to optical phonons, which is forbidden by energy conversation (remember that the optical phonon energy is $34.4$~meV and mostly excitons with $\epsilon_x\approx 0~$meV are excited). Note that the finite value of $N^{\text{cap}}$ at $t=0$ is due to the excitation condition, which excites a small fraction of localized excitons. The small increase in occupation of the bound states can be traced back to a very small population of a weakly bound state at $\epsilon_x=-2~$meV by acoustic phonons, which results in a weak additional dephasing of the coherent occupation.

In the second example we take a potential depth of $V_0=60~$meV (red curves in Fig.~\ref{Fig_coh_dephase}), which has the lowest bound state at $\epsilon_1=-32~$meV. Here, scattering due to optical phonons is possible. Indeed we observe an additional decrease in coherent population of about $19\%$ on a much slower timescale than the initial build-up of the polaron. The additional decrease is linked to the capture of excitons into localized states as seen in Fig.~\ref{Fig_coh_dephase}(b), where the drop in $N_{\mathrm{coh}}$ corresponds to a rising $N^{\mathrm{cap}}$. This shows that the captured excitons are almost completely incoherent. In other words, during the scattering into the localized potential the coherent excitons become incoherent. This is confirmed by looking at the fraction of captured coherent excitons $N^{\text{cap}}_{\text{coh}}$ shown as a dotted line in Fig.~\ref{Fig_coh_dephase}(b), which is very small for both potentials. 

Another feature of the capture is that the trapped excitons perform an oscillation with a periode of about $0.8~$ps ontop of a gradual increase (see inset in Fig.~\ref{Fig_coh_dephase}(b)), which is most prominent when only optical phonons are considered (dashed lines). This is similar to the dynamics as previously found for capture from a one-dimensional to a zero-dimensional system \cite{Glanemann05} and to the dynamics in a two-level system. The analogy of the two-level system is actually applicable, because closer inspection reveals that the capture results dominantly from one very weakly localized state with $\epsilon_x=-0.4~$meV which is coupled to the localized state at $\epsilon_1=-32~$meV, while all other optically excited states do not effectively couple to the strongly localized state. 

Interestingly, the inclusion of acoustic phonon emission prevents the release of captured carriers, thereby damping the oscillation [solid lines in Fig.~\ref{Fig_coh_dephase}(b)]. We note that a direct population of the state with $\epsilon_x=-32~$meV by acoustic phonons is essentially impossible due to the small energy of acoustic phonons. Only the combination of both phonon branches leads to the almost monotonically increasing capture. This can be interpreted again within the analogy of the two-level system weakly coupled with a dispersionless bosonic mode given by the optical phonons, which results in Rabi oscillations. The coupling leads to an energy splitting, which in our case is of about $5~$meV deduced from the oscillation period of $0.8$~ps, such that acoustic phonons can lead to scattering between these split states thereby damping the Rabi oscillations \cite{Groll20}. This is confirmed by looking at longer simulation times, as depicted in the inset of Fig.~\ref{Fig_coh_dephase}(b), where the dynamics up to $2$~ps is plotted and a considerable damping of the oscillation can be seen when including acoustic phonons. A similar relaxation like this as a two-phonon process is known to be efficient in, e.g., carrier relaxation within a quantum dot \cite{Inoshita92}. Note that in a two-level system coupled to a dispersionless mode no real relaxation to the lower state can occur and only Rabi oscillations would be seen. Here, however, the gradual increase in captured density stems from coupling of the weakly localized state to delocalized states \cite{Seebeck05,Muljarov07}.

\subsection{Influence of incoherent exciton scattering}
Now we want to focus on the influence of scattering between incoherent excitons on the capture dynamics. In the previous section, we have already seen that due to the polaron formation incoherent excitons are created independently of the localization potential (cf. Fig.~\ref{Fig_coh_dephase}).  In particular, we want to examine if the scattering from the delocalized incoherent excitons is also an efficient path for capture. 

The dynamics of incoherent excitons are of crucial importance for optical experiments and optoelectronic applications, where incoherent excitons are usually the main source of luminescence \cite{Thraenhardt00,Rossi02,Brem18}. While the coherent polarization $P_x$ is finite only for energies around $\epsilon_x=0~$meV, the incoherent excitons are generated at high kinetic energies to compensate the negative interaction energy due to the formation of the polaron (even at $0~$K). In the simulations a numerical cutoff of $\epsilon_x=100~$meV was necessary for converged results. We emphasize that coherent excitons recombine resulting in spontaneous emission, which usually leads to a very fast decay of coherent excitons \cite{Moody15,Selig16}. This effect plays a major role for delocalized coherent excitons after spatially broad excitations and is neglected in this work, while incoherent excitons usually decay on a much longer timescale due to scattering outside the light cone \cite{Selig18}. Additionally we want to stress that the dominant capture in Fig.~\ref{Fig_coh_dephase} happened due to transitions from a very weakly localized state to a strongly localized state. Transitions from delocalized states with $\epsilon_x\ge 0~$meV are found to be very inefficient, making the study of incoherent exciton capture even more important.

\begin{figure}[t!]
\includegraphics[width=\columnwidth]{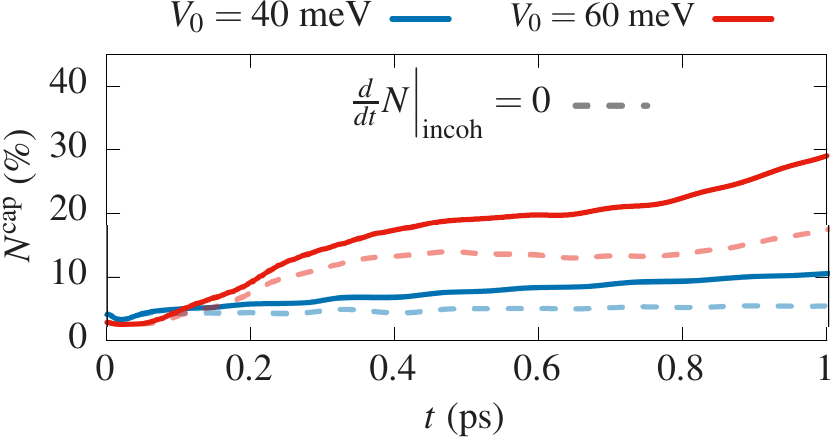}
\caption{Fraction of excitons in localized states $N_{\mathrm{cap}}$ calculated with (solid lines) and without (dashed lines) scattering between incoherent excitons for $V_0=40$~meV (blue curves) and $V_0=60$~meV (red curves).}
\label{Fig_incoh_cap}
\end{figure}

To focus on the influence of scattering of incoherent excitons on the capture process, we display in Fig.~\ref{Fig_incoh_cap} the fraction of excitons in localized states $N_{\mathrm{cap}}$ including now scattering between incoherent excitons. As reference point, we also include the curve, where this is switched off ($\left.\frac{d}{dt}N_{xx'}\right|_{\mathrm{incoh}}=0$, corresponding to the former section).  

For the potential with $V_0=40~$meV, we find that now the captured density increases from about $5$\% to about $10$\%. Here all captured excitons stem from incoherent excitons which form due to the build-up of the polaron. Without scattering between incoherent excitons the captured density stays approximately the same over the whole simulation duration. This underlines the importance of the polaron at low temperatures, because it leads to incoherent excitons with a broad energetic distribution, therefore leading to exciton capture quite independent of the specific energy of the localized state.

For the potential with $V_0=60~$meV, the final amount of captured carriers increases from about $17$\% to about $29$\% by considering scattering between incoherent excitons. Also here we find that the influence of the scattering of incoherent excitons has a significant influence on the capture rate.

\subsection{Spatiotemporal dynamics}

\begin{figure}[t!]
\includegraphics[width=\columnwidth]{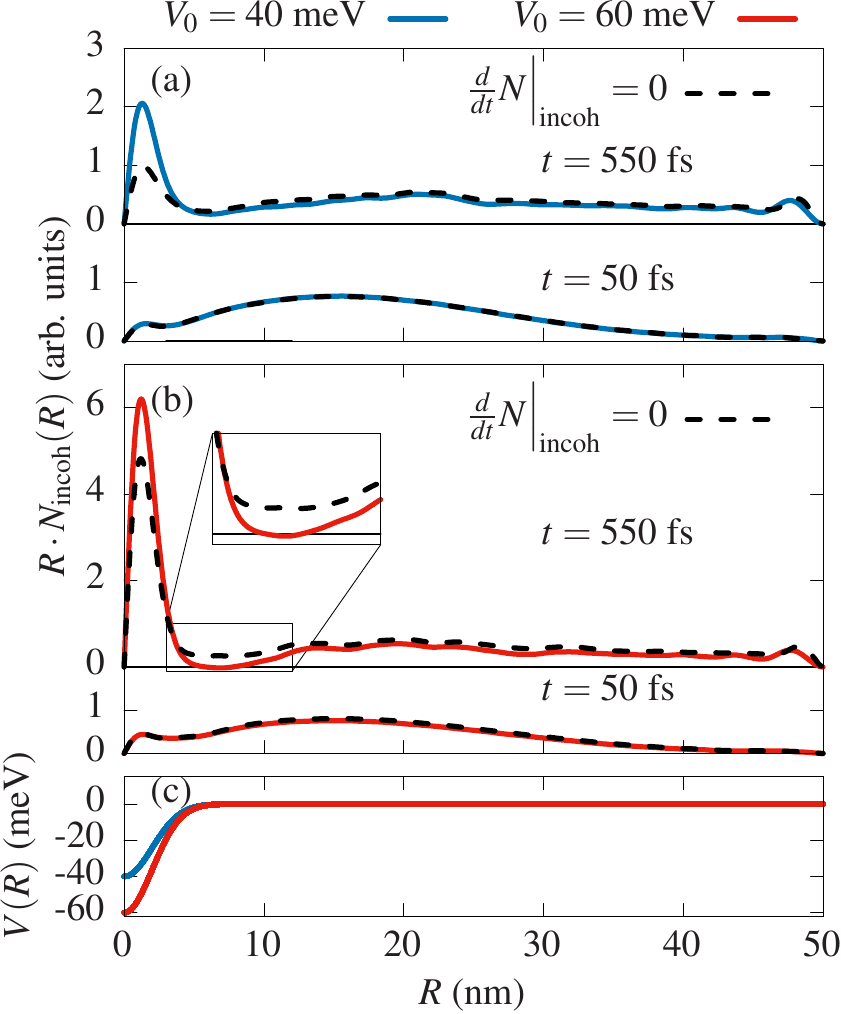}
\caption{Spatial density of incoherent excitons at two snapshots at $t=50~$fs and $t=550~$fs with (solid lines) and without (dashed lines) scattering between incoherent excitons for (a) $V_0=40$~meV (blue curves) and (b) $V_0=60$~meV (red curves). (c) Plot of the potentials.}
\label{Fig_incoh_dyn}
\end{figure}

Within our non-diagonal treatment of the density matrix we calculate the spatiotemporal dynamics for this inhomogeneous problem providing a picture of the capture dynamics in real space. As we have seen that the captured excitons are mainly incoherent, we plot the spatial density of the incoherent excitons $N_{\mathrm{incoh}}(R)$  in Fig.~\ref{Fig_incoh_dyn} for (a) $V_0=40$~meV and for (b) $V_0=60$~meV. This quantity is given by 
$$N_{\mathrm{incoh}}(\mathbf{R})=\sum\limits_{x,x'} \Psi_{x}^*(\mathbf{R})\Psi_{x'}(\mathbf{R}) N_{xx'}$$
and is thereby directly linked to the non-diagonal elements of the incoherent part of the density matrix.
Due to the radial symmetry $N_{\mathrm{incoh}}(\mathbf{R})$ only depends on $R=|\mathbf{R}|$ and we multiply the density with $R$, such that the area under the curve is a measure for the total number of incoherent excitons. Note that we used a disk of a $50~$nm radius as our simulation domain, such that the density vanishes at $R=50~$nm in any case (see App.~\ref{App_symmetry}).

At early times the spatial density of the incoherent excitons builds up from coherent exciton density. Accordingly, the incoherent density follows the spatial profile of coherent excitons and we see at $t=50~$fs a Gaussian profile in space (remember that we multiplied the density with $R$, such that at $R=0$ a node occurs). Here, we have mostly delocalized states, while due to the excitation already a small density at the potential is visible, though no considerable capture took place [cf. Fig.~\ref{Fig_coh_dephase}(b)].

After some time, scattering into the localized states takes place and accordingly, we see that the spatial density increases in the region of the localized potential plotted in Fig.~\ref{Fig_incoh_dyn}(c). The capture dynamics is very similar for both potentials of $V_0=40$ and $60$~meV. At the edge of the simulation box, i.e. close to $R=50$~nm, one can see some spatial oscillations, which is a numerical effect due to the wave packet interfering with the reflected one at the boundary. This, however, does not affect the capture dynamics since the localized states are far away from the simulation boundary. The spatiotemporal dynamics shows, that the capture happens only locally, in consistency with previous findings on the locality of capture processes in systems of lower dimensionality \cite{Glanemann05,Reiter07,Rosati17,Lengers17}. Further capture into the localized state occurs when density from the outer regions of the sample travel onto the potential leading to the gradual increase of captured excitons in Fig.~\ref{Fig_incoh_cap} for both potentials

It is interesting to have a closer look at the carrier capture for $V_0=60~$meV. Here, the capture process is so efficient that after about $550~$ fs no incoherent density is left in the direct vicinity outside the QD potential, as shown in the inset of Fig.~\ref{Fig_incoh_dyn}(b). By comparing the curves without (dashed lines) and with (solid lines) scattering between incoherent excitons, we see that this strong capture is only possible by the latter mechanism. We further observe a small negativity in the density, which is a well-known effect in 2BA. More details on the negativities can be found in App.~\ref{App_negativities}.

We note that the high capture efficiency found in our calculations is consistent with the experimental findings for TMDCs that in spatially resolved photoluminescence maps the regions in the vicinity of localized excitons show strongly reduced emission from free exciton states \cite{Tonndorf15}.

\section{Capture from purely incoherent excitons}
Having illustrated the importance of exciton capture from incoherent excitons, we will discuss this capture process in more detail by directly assuming an initial condition for a purely incoherent exciton distribution. This situation mimics an off-resonant excitation as in photoluminescence experiments, where first the excitons relax to quasi-free $1s$ excitons and are then trapped \cite{Tonndorf15,Srivastava15,Feierabend19}. We approximate this process by assuming that after optical excitation the excitons first relax to a quasi-stationary distribution within the delocalized states of the $1s$ subspace and then capture processes set in. 

We give the initial condition of incoherent excitons in terms of the Wigner function 
\[
	N(\mathbf{K},\mathbf{R})=\sum\limits_{\mathbf{Q}}\delta\left\langle\cre{P}_{\mathbf{K+\frac{Q}{2}}}\hat{P}_{\mathbf{K-\frac{Q}{2}}}\right\rangle e^{i\mathbf{Q\cdot R}}
\]
with
	\begin{align}
	N(\mathbf{K},\mathbf{R})=N_0\exp\left(-\frac{\left(\frac{\hbar^2K^2}{2M}-E_0\right)^2}{2\sigma_E^2}\right)\exp\left(-\frac{R^2}{2\sigma_R^2}\right) 
	 \notag 
	\end{align}
with an energetic broadening $\sigma_E$, a spatial width $\sigma_R$ and a central energy $E_0$. We take for the spatial width $\sigma_R=15~$nm and will vary $E_0$ and $\sigma_E$. This value of $\sigma_R$ results in the same spatial width of the initial exciton density as the optical pulse considered in the former section because $u(\mathbf{R})^2=\exp\left(-R^2/(2\sigma_R^2)\right)$ holds. After initialization, we transform the Wigner function into the eigenbasis with the disk of $50~$nm obtaining $N_{xx'}$, while omitting all contributions where $x,x'$ are localized states. Then we propagate the incoherent density using the equations of motion derived in Sec.~\ref{sec:theory}. 

\subsection{Capture dynamics}
 
\begin{figure}[t!]
\includegraphics[width=\columnwidth]{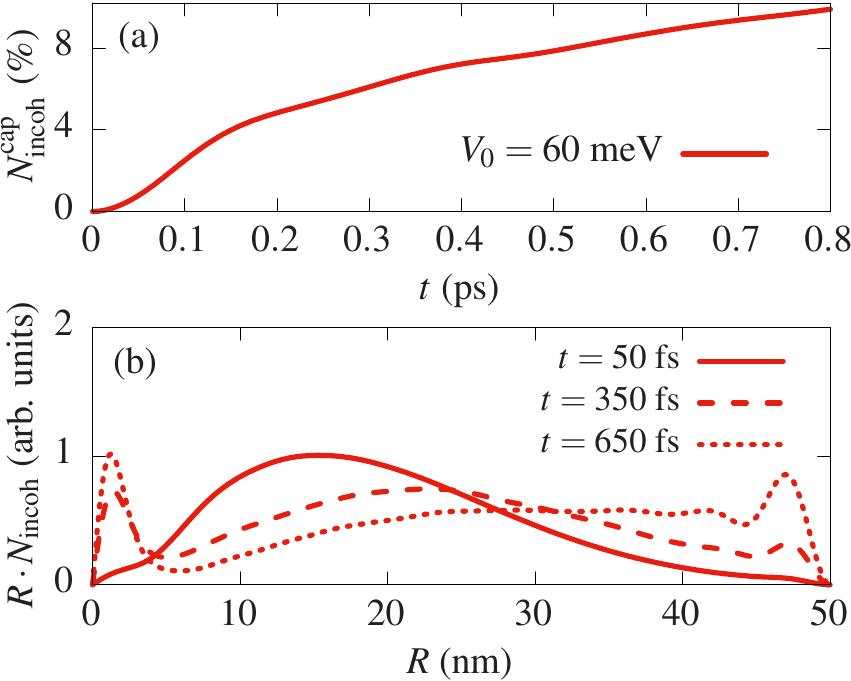}
\caption{Exciton capture from an incoherent distribution for $V_0=60$~meV. (a) Fraction of captured excitons. (b) Spatial dynamics of the incoherent exciton density at three snapshots at $t=50~$fs, $t=350~$fs and $t=650~$fs.}
\label{Fig_incoh_WP_cap}
\end{figure}

The capture dynamics from an incoherent initial condition is shown in Fig.~\ref{Fig_incoh_WP_cap}(a), where we assumed $E_0=0~$meV and  $\sigma_E=10~$meV and considered the potential with $V_0=60$~meV. We see that the fraction of captured excitons shows a monotonic increase, which is almost linear after about $0.2~$ps. The linear increase can be traced back to the broad spatial distribution of excitons, because excitons travel continuously into the vicinity of the potential where they can be captured \cite{Reiter07}.

The corresponding spatiotemporal dynamics is displayed in  Fig.~\ref{Fig_incoh_WP_cap}(b), where we show $R N_{\text{incoh}}$ for $t=50,~350$ and $650$~fs. Also here, we  observe a gradual capture of excitons from the vicinity of the localized potential leading to a depletion of delocalized excitons as in the previous section.
\subsection{Evaluation of the resonance condition}

We will now turn to the question what determines the efficiency of the capture process. We have already seen that optical phonons provide the most efficient relaxation channel. However, due to their constant energy, from a semiclassical point of view an efficient capture can take place only for matching energy conditions (i.e., when the delocalized states are about one optical phonon energy above the bound states). This is analogous to the phonon bottleneck in quantum dots where strongly suppressed relaxation between discrete states was expected and measured \cite{Heitz01,Urayama01}. In conventional semiconductors it was shown in quantum kinetic calculations that the presence of the polaron, especially within the continuum of states, in combination with the intrinsic local nature of capture strongly broadens the energy selection rule \cite{Seebeck05,Glanemann05,Muljarov07}. While such an energetic broadening can be incorporated in simplified models to enable simulations in systems of low symmetry \cite{Rosati17,Rosati19}, its quantitative value is of quantum kinetic nature. Therefore, it is interesting to study this effect in TMDCs and quantify the polaron-related energy broadening in TMDCs. 

\begin{figure}[t!]
\includegraphics[width=\columnwidth]{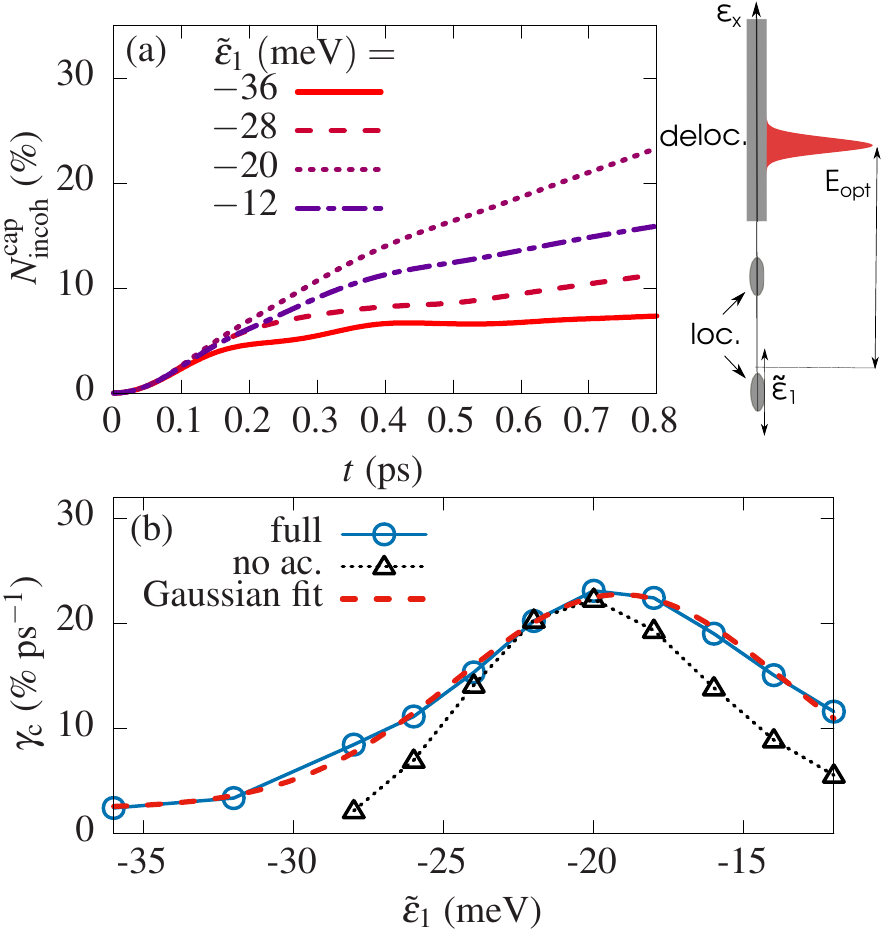}
\caption{(a) Captured density as function of time for different values of the lowest energy $\tilde{\epsilon}_1$. The setup is sketched on the right, where the excitonic states are marked as grey areas and the initialized distribution as a red gaussian. (b) Capture rate $\gamma_c$ for the full calculation (blue dots) and without acoustic phonons (black triangles). The red dashed line is a Gaussian fit.}
\label{Fig_capture_final}
\end{figure}

For our simulations we take the potential with $V_0=60$~meV, which has two localized states at $\epsilon_1=-32~$meV and $\epsilon_2=-10~$meV. We then adjust the detuning between the energetically lowest localized state and the scattering continuum by artificially shifting the energy of the lowest localized state by an amount $\delta$ to $\tilde{\epsilon}_1=\epsilon_1+\delta$, while leaving all other quantities unchanged. We also use the unchanged energy $\epsilon_1$ for the calculation of the damping rate $\Gamma_x$. A sketch of our setup is given on the right side of Fig.~\ref{Fig_capture_final}(a). We choose an initial distribution of incoherent excitons with the central energy $E_0=10~$meV and a small energetic broadening of $\sigma_E=2~$meV. Because the energetic distribution is very narrow, energy selection rules should result in a sharp resonance of the capture efficiency at $\tilde{\epsilon}_1=E_0-\hbar\omega_{\mathrm{op}}\approx-24~$meV. 

In Fig.~\ref{Fig_capture_final}(a) we plot the fraction of localized excitons $N^{\mathrm{cap}}_{\mathrm{incoh}} = (N_{11}+2N_{22})/N_{\mathrm{incoh}}$ for different $\tilde{\epsilon}_1$ as function of time. Note that the state $2$ stems from angular momenta $l=\pm 1$ and therefore has to be counted twice. For short times up to about $0.2$~ps the capture process is very similar for all $\tilde{\epsilon}_1$ due to energy-time uncertainty. After about $0.4~$ps one can observe a linear increase strongly dependent on $\tilde{\epsilon}_1$ leading to different final values of  $N^{\mathrm{cap}}_{\mathrm{incoh}}$. We quantify the amount of captured carriers by the capture rate $\gamma_c$, which is obtained by a linear fit of the dynamics for times larger than $0.4~$ps and summarize the results in Fig.~\ref{Fig_capture_final}(b).

\begin{figure} [t!]
\includegraphics[width=\columnwidth]{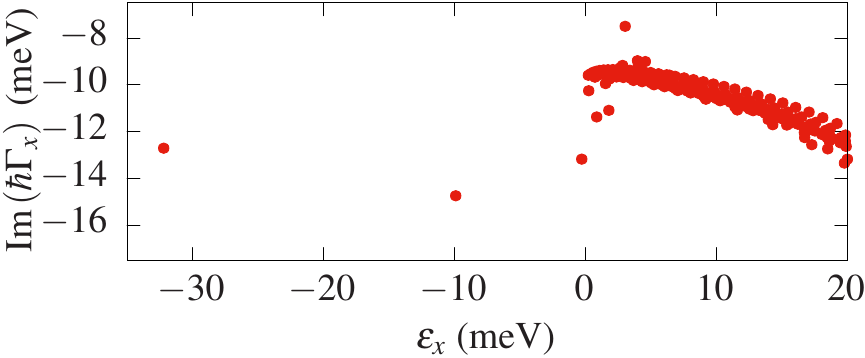}
\caption{Calculated energy renormalizations $\mathrm{Im}(\hbar\Gamma_x)$ for the state $x=(n,l)$.}
\label{Fig_polaronshifts}
\end{figure}

The capture rate as function of $\tilde{\epsilon}_1$ shows a clear resonance behavior of the capture efficiency. However, the maximum capture rate is obtained for $\tilde{\epsilon}_1 \approx -20$~meV, which is significantly higher than the expected resonance at $\tilde{\epsilon}_1=-24~$meV. To understand this, we have to consider that transitions do not take place with respect to the single-particle energies, but with respect to polaron-shifted energies. These energy renormalizations are considered by $\mathrm{Im}\left(\hbar\Gamma_x\right)$ in our calculations, which can be seen by performing the completed collision limit resulting in
	\begin{align}
	\lim\limits_{T_{\mathrm{av}}\rightarrow\infty}\mathrm{Im}\left(\hbar\Gamma_x\right)
	=&\sum\limits_{\mathbf{Q},x',i} |g_{\mathbf{Q},i}|^2\mathcal{P}\left(\frac{|K_{x\mathbf{Q}x'}|^2}{\epsilon_{x}-\epsilon_{x'}-\hbar\omega_{\mathbf{Q},i}}\right)\nonumber\\
	\hat{=}&\Delta\epsilon_x^{\mathrm{pol}} ,
	\label{Eq_pol_shift}
	\end{align}
where the latter expression is identified as the polaron shift $\Delta\epsilon_x^{\mathrm{pol}}$ calculated in second order stationary perturbation theory with the Cauchy principal value $\mathcal{P}$.
The calculated renormalizations $\mathrm{Im}\left(\hbar\Gamma_x\right)$ for the lowest states are illustrated in Fig.~\ref{Fig_polaronshifts}. Remarkably, the polaron shift depends sensibly on the considered state. For the bound state at $\epsilon_1=-32$~meV, we find a polaron shift of about $-12.5$~meV, while for the delocalized states around $\epsilon_x\approx 10$~meV,the shifts are about $-10~$meV. Accordingly, the difference in polaron shifts is about $2-3$~meV and brings the resonance seen in Fig.~\ref{Fig_capture_final}(b) closer to its expectation. We additionally performed simulations with an artificially decreased exciton-phonon interaction for optical phonons indeed resulting in a resonance at $\tilde{\epsilon}_1=-24~$meV (not shown here). We conclude that in TMDCs the exciton-phonon coupling is so strong that resonance-conditions as derived from unperturbed single-particle energies may be misleading.

Remarkably, the resonance found in Fig.~\ref{Fig_capture_final}(b) exhibits a strong broadening of the resonance. This demonstrates that there is no phonon bottleneck present even for an energetically very narrow distribution within the delocalized states. To extract the width of the resonance we fit the curve by a Gaussian [dashed red line in Fig.~\ref{Fig_capture_final}(c)]
\[
	G(\tilde{\epsilon}_1)=G_0+A\exp\left(-\frac{(\tilde{\epsilon}_1-\tilde{\epsilon}_0)^2}{2\sigma_{\epsilon}^2}\right).
\]
From this we extract a broadening of $\sigma_{\epsilon}=5.4~$meV.  Assuming that the observed capture results from a convolution of the capture rate with the initialized exciton density with a width of $2~$meV, we compute an energetic broadening of the capture $\bar{\epsilon}=\sqrt{\sigma_{\epsilon}^2-(2~\mathrm{meV})^2}=5~$meV showing a strong broadening of about $15\%$ of the optical phonon energy. We note that the broadening seen in the above results is not solely related to the damping $\Gamma_x$ introduced in the theory section which transforms the $\delta$-function selection rule of a semiclassical theory to a Lorentzian selection rule, but is mainly a feature of the quantum kinetic treatment.
\begin{figure}[t]
\includegraphics[width=\columnwidth]{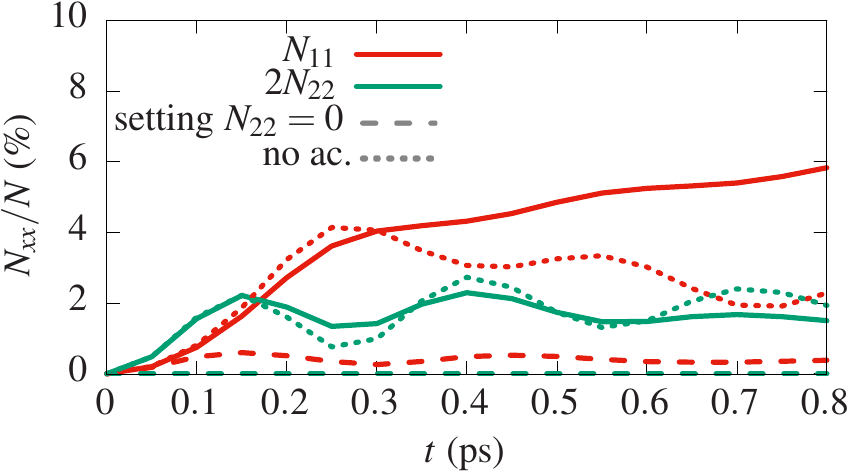}
\caption{Incoherent exciton occupations for the two lowest localized states in the case of $\tilde{\epsilon}_1=-36~$meV shown in solid lines, the same computation with $N_{22}=0$ shown as dashed lines and the calculation without acoustic phonons shown as dotted lines.}
\label{Fig_2Phonon}
\end{figure}

Finally, we want to consider the offset of the capture efficiency with respect to the resonance, leading to a finite capture rate  even for $\tilde{\epsilon}_1=-36~$meV. To understand the offset, we take a look at the dynamics of the localized states $1$ and $2$ shown in Fig.~\ref{Fig_2Phonon} for the case of $\tilde{\epsilon}_1=-36~$meV. The dynamics of $N_{11}$ and $N_{22}$ show an oscillation with a phase shift of $\pi$ which is a result of the coupling by optical phonons leading to Rabi oscillations between two discrete energy states. 

We can now artificially turn off the capture into state 2, such that $N_{22}$ stays zero (dashed lines). Then the occupation of the lowest bound state $N_{11}$ stays very small over the whole simulation range. This leads to the interpretation that a strong capture into state $1$ takes place by the 2-step process via state 2. We additionally plot the results without acoustic phonons as dotted lines showing that the inclusion of acoustic phonons again leads to a possible relaxation from state $2$ to $1$ by the same reasoning as in Sec.~\ref{Sec_coh_dephase} and thereby to the finite capture rate in Fig.~\ref{Fig_capture_final}(b) for the highly off-resonant case. Without acoustic phonons we only observe a small increase of $N_{22}$ ontop of the oscillations because relaxation from the continuum of states is possible by optical phonons.

When neglecting the influence of acoustic phonons on the capture, this results in a strong reduction of the offset and also in a reduced width of $\gamma_c$ as shown in Fig.~\ref{Fig_capture_final}(c) (black triangles) which can be fitted by $\sigma_{\epsilon}=4.2~$meV resulting in $\bar{\epsilon}\approx 3.5$~meV, being $1.5~$meV smaller than in the case with acoustic phonons. For highly off-resonant capture scenarios acoustic phonons are able to channel the capture process therefore effectively broadening the capture resonance. Note that a linear fit after $0.4~$ps in the case without acoustic phonons is not reasonable for $\tilde{\epsilon}_1<-28~$meV because the dynamics of $N_{\mathrm{cap}}$ resembles oscillatory dynamics rather than a linear dependence on time. In any case the absolute values of $N^{\mathrm{cap}}$ without acoustic phonons are considerably lower in this range of $\tilde{\epsilon}_1$.

\section{Conclusion}
In conclusion we have presented a quantum kinetic treatment of exciton capture due to phonons in TMDCs. In the derivation of the quantum kinetic equations of motion using a non-diagonal density matrix for this inhomogeneous problem we treated the dynamics within the delocalized states by making use of well-established theories,i.e. , the TCL approach and a Boltzmann-like equation. For the dynamics which involve localized states we used a damped second order Born approximation to truncate the hierarchy of equations of motion. 

By distinguishing between coherent and incoherent excitons, we have demonstrated that the capture process results in incoherent excitons and that scattering between incoherent excitons plays a major role. The analysis of the interplay between optical and acoustic phonons showed that for an efficient capture on the one hand an optical phonon is emitted and on the other hand acoustic phonons prevent the release of the captured carriers resulting in a monotonically increasing capture. Additionally, we discussed the spatiotemporal capture dynamics, underlining the importance of scattering of incoherent excitons with phonons for the carrier capture.

Focussing on a purely incoherent density, we furthermore showed that the resonance condition for capture by optical phonon emission is considerably broadened in quantum kinetic calculations. We related the resonance condition of the capture rate to the strong polaron shifts, which are different for every excitonic state. Additionally, we discussed that an ultrafast relaxation mediated by two subsequent phonon-emission processes including acoustic and optical phonons becomes possible. 

Our work sheds new light on the impact of exciton-phonon interaction on the ultrafast spatiotemporal dynamics in TMDC monolayers, especially concerning localized excitons which are at the heart of TMDC applications in quantum information technology.

\acknowledgements
 F.~L. and D.~E.~R. acknowledge financial support by the Deutsche Forschungsgemeinschaft (DFG) by the project
406251889 (RE 4183/2-1).

\appendix

\section{Bound states and equations of motion for radial symmetry}
\label{App_symmetry}
The bound excitonic states are calculated in real space by solving
	\begin{align*}
	\left[-\frac{\Delta_{\mathbf{r}}}{2M}+{V}_{C,1s}(\mathbf{R})\right]{\Psi}_x(\mathbf{R})=\epsilon_x{\Psi}_x(\mathbf{R}).
	\end{align*}
with the exciton mass $M=m_e+m_h$ composed of the effective mass of electron $m_e$ and hole $m_h$.
For the solution we consider a disk of radius $R_{\mathrm{max}}=50~$nm and solve with Dirichlet boundary conditions with ${\Psi}_x(|\mathbf{R}|=R_{\mathrm{max}})=0$. Since the free solution without additional confinement $V_{C,1s}$ is given by Bessel functions of the first kind, we expand the wavefunction in Bessel functions which is a good choice for spherical symmetric problems in which ${\Psi}_x(\mathbf{R})={\Psi}_{n,l}(R)e^{il\phi}$ can be imposed with radial quantum number $n$, angular quantum number $l$ and $\mathbf{R}=(R\cos(\phi),R\sin(\phi))$.

While the equations of motion in the main text are given for an arbitrary basis, the occuring summations can be greatly simplified if the rotational symmetry of the considered states is used. 
For calculation of the form factors $K_{x,\mathbf{Q},x'}$ it is instructive to use the Jacobi-Anger identity $e^{ix\cos(\phi)}=\sum\limits_{n=-\infty}^{\infty} i^nJ_{n}(x)e^{in\phi}$ with the $n$-th Bessel function $J_n$ to arrive at
	$$K_{x\mathbf{Q}x'}= i^{l-l'}K_{(n,l),Q,(n',l')}e^{-i(l-l')\phi_Q}$$
with $\mathbf{Q}=(Q\cos(\phi_Q),Q\sin(\phi_Q))$, $x=(n,l)$ and 
	$$K_{(n,l),Q,(n',l')}=2\pi\int\limits_0^{R_{\mathrm{max}}} {\Psi}_{(n,l)}^*(R){\Psi}_{(n',l')}(R)J_{l-l'}(QR) RdR.$$
	
This allows one to redefine the dynamical variables, e.g.,
	$$S_{n'l',i}^{(P,+)}(Q):=\frac{i^{-l'}}{2\pi}\int_0^{2\pi}S_{n'l',i}^{(P,+)}(\mathbf{Q})e^{il'\phi_Q}d\phi_Q$$
in Eq.~(\ref{eom_S_p}), because only $l=0$ polarizations $P_{(n,l)}$ are excited by a Gaussian beam without orbital angular momentum. Since no quantity in the equation of motion for $S^{(P,+)}$ depends on $\phi_Q$ other than $\left(K_{(n,l),Q,(n',l')}\right)^*e^{i(l-l')\phi_Q}$, the integration over $\phi_Q$ can be readily performed leading to the restriction $l=0$ in this case. This also shows that after coherent excitation of the $l=0$ coherence, no coherence of angular momentum $l\neq 0$ is excited. We can therefore omit one summation, which leads to the equation of motion 
\begin{align*}
i\hbar\frac{d}{dt}S_{n'l',i}^{(P,+)}(Q)&=
\left(\epsilon_{x'}^{\lambda}-\hbar\omega_{Q,i}\right)S_{n'l',i}^{(P,+)}(Q)\\
&+n_{Q,i}\sum\limits_{n}|g_{Q,i}|^2\left(K_{(n,0),Q,(n',l')}\right)^* P_{n0}\nonumber.
\end{align*}
Here we made use of $\hbar\omega_{\mathbf{Q},i}=\hbar\omega_{Q,i}$ and $g_{\mathbf{Q},i}=g_{Q,i}$ in the deformation potential approximation used here. 
Thereby the $\phi_Q$-integration and one summation over angular momenta is eliminated. These redefinitions of dynamical variables can be performed for all other quantities analogously. The most important implications of the used symmetry are $N_{xx'}=N_{(nl)(n'l')}\delta_{l,l'}$ and $P_x=P_{nl}\delta_{l,0}$. This redefinition of variables is independent of their specific treatment, i.e., it applies to the dynamics within the delocalized states ($\mathbf{DL}$) and localized states ($\mathbf{L}$).

\section{Damping due to higher-order correlations}
\label{App_gamma}
We here perform the derivation of the damping rate $\Gamma_x$ as introduced in Eq.~(\ref{eom_coh_pol_L}).
Considering higher order correlations in fourth Born approximation (4BA) one arrives at an additonal contribution for the phonon-assisted polarization reading
$$i\hbar\frac{d}{dt}S_{x',i}^{(P,-)}(\mathbf{Q})=...+\sum\limits_{x,\mathbf{Q}',j} g_{\mathbf{Q},i}g_{\mathbf{Q'},j}K_{x'\mathbf{Q'}x}\delta\langle \hat{P}_{x}\hat{b}_{\mathbf{Q},i}\hat{b}_{\mathbf{Q}',j}\rangle$$
where we concentrate on $T=0$~K for now. In that case the equation of motion for the two-phonon-assisted correlation reads
\begin{align*}
&i\hbar\frac{d}{dt}\delta\langle \hat{P}_{x}\hat{b}_{\mathbf{Q},i}\hat{b}_{\mathbf{Q}',j}\rangle
\\
&=(\epsilon_x+\hbar\omega_{\mathbf{Q},i}+\hbar\omega_{\mathbf{Q}',j})\delta\langle \hat{P}_{x}\hat{b}_{\mathbf{Q},i}\hat{b}_{\mathbf{Q}',j}\rangle\\
&+\sum\limits_{x''}\left(\frac{g_{\mathbf{Q},i}^*}{g_{\mathbf{Q'},j}}(K_{x''\mathbf{Q}x})^*S_{x'',j}^{(P,-)}(\mathbf{Q}')\right)\\
&+\sum\limits_{x''}\left(\frac{g_{\mathbf{Q}',j}^*}{g_{\mathbf{Q},i}}(K_{x''\mathbf{Q}'x})^*S_{x'',i}^{(P,-)}(\mathbf{Q})\right).
\end{align*}
This correlation is now adiabatically integrated in a Born approximation such that one can include its contribution approximately without treating it as a dynamical variable. This reads
\begin{align}
&\delta\langle \hat{P}_{x}\hat{b}_{\mathbf{Q},i}\hat{b}_{\mathbf{Q}',j}\rangle=
\label{eom_2phonon_assisted_appendix}\\
&-\frac{i}{\hbar}\sum\limits_{x''} \frac{g_{\mathbf{Q},i}^*}{g_{\mathbf{Q'},j}}(K_{x''\mathbf{Q}x})^*f_{x,x'',\mathbf{Q},i}(t)S_{x'',j}^{(P,-)}(\mathbf{Q}')\nonumber \\
&-\frac{i}{\hbar}\sum\limits_{x''}\frac{g_{\mathbf{Q'},j}^*}{g_{\mathbf{Q},i}}(K_{x''\mathbf{Q}'x})^*f_{x,x'',\mathbf{Q}',j}(t)S_{x'',i}^{(P,-)}(\mathbf{Q})\nonumber
\end{align}
with the introduced function 
	$$f_{x,x'',\mathbf{Q},i}(t)=\int_0^t \exp\left(-\frac{i}{\hbar}\left(\epsilon_x-\epsilon_{x''}+\hbar\omega_{\mathbf{Q},i}\right)\tau\right)d\tau$$
	where we chose $t_0=0$ as the time of optical excitation by a $\delta$-pulse. Plugging this expression for the two-phonon-assisted correlation into the equation of motion for $S_{x'}^{(P,-)}(\mathbf{Q},i)$ shows that the $\mathbf{Q}'$ summation does not contain dynamical variables for the second term on the right hand side of Eq.~\eqref{eom_2phonon_assisted_appendix}. Performing a random phase approximation, the first term on the right hand side of Eq.~\eqref{eom_2phonon_assisted_appendix} can be neglected. We therefore only consider the second term on the right hand side where we additionally only take into account the diagonal term for $x''=x'$ in the equation of motion for $S_{x'}^{(P,-)}(\mathbf{Q},i)$ being the only remaining term in the homogeneous limit leading to strong improvement of the dephasing dynamics in Ref.~\cite{Lengers20}. These contributions then result in a damping rate for the correlations according to
		$$i\hbar\frac{d}{dt}S_{x',i}^{(P,-)}(\mathbf{Q})=...-i\hbar\gamma_{x'}S_{x',i}^{(P,-)}(\mathbf{Q})$$
with the damping rate as introduced in the theory section.

\section{Influence of the damping}
\label{App_negativities}
\begin{figure}[h!]
\includegraphics[width=\columnwidth]{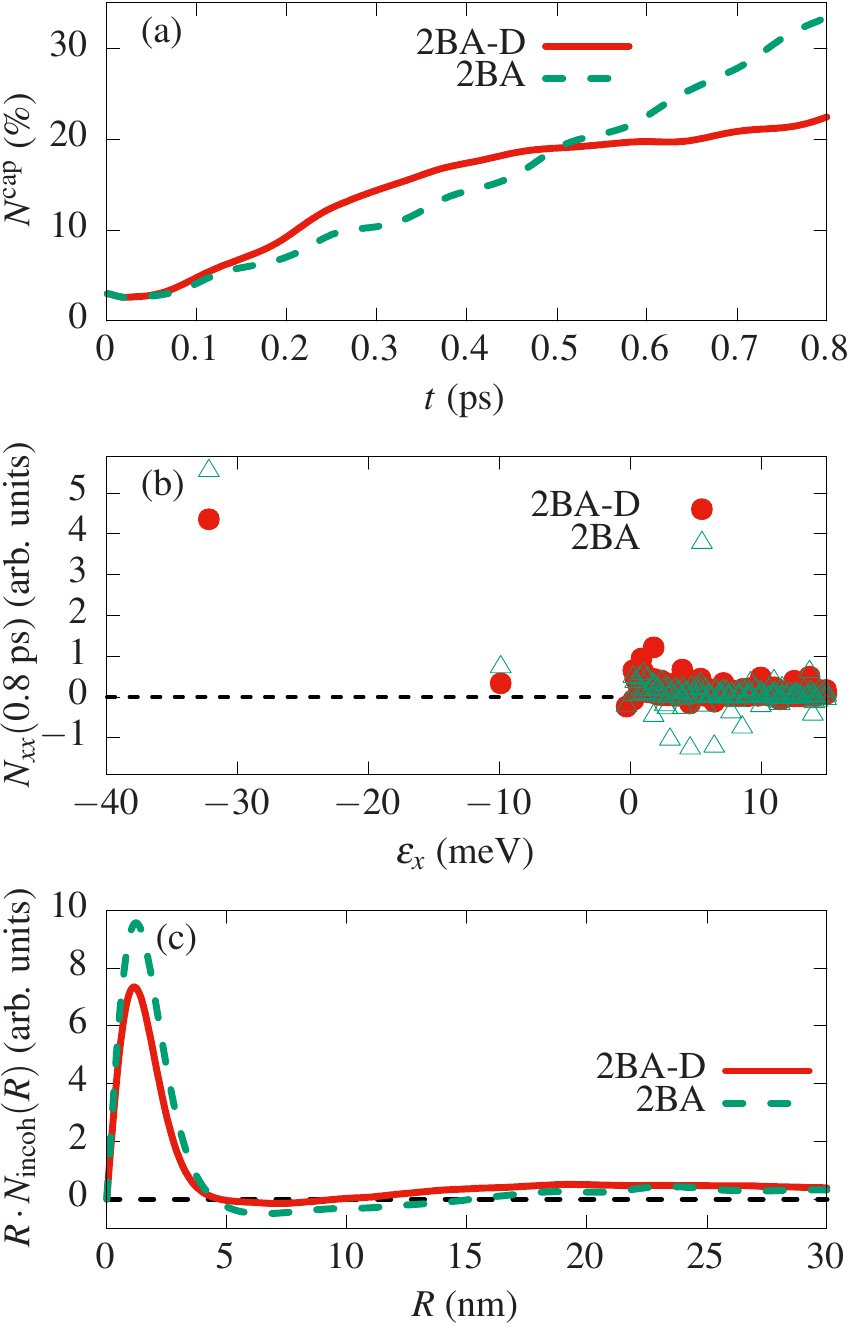}
\caption{Comparison between 2BA and 2BA-D. (a) Fraction of captured exciton density as in Fig.~\ref{Fig_incoh_cap}(a) with (solid red line) and without (dashed green line) damping. (b) Occupation of incoherent excitons after $0.8$~ps with (red dots) and without (green triangles) damping. (c) spatial density of incoherent excitons after $0.8$~ps with (red solid line) and without (green dashed line) damping.}
\label{Fig_2D0D_0K}
\end{figure}

Below we compare two calculations with (2BA-D) and without (2BA) the introduced damping rates $\Gamma_x$ in Fig.~\ref{Fig_2D0D_0K}. We excite with a $\Delta_E=30~$nm focussed $\delta$-pulse above a $\Delta_R=2~$nm wide and $V_0=60~$meV deep QD potential and observe subsequent capture of carriers. This corresponds to the dynamics shown in Fig.~\ref{Fig_incoh_cap} and Fig.~\ref{Fig_incoh_dyn}(b).
One observes in Fig.~\ref{Fig_2D0D_0K}(a) that the total amount of captured excitons does not change much up to $0.6~$ps, but that the final captured density is lower with damping in 2BA-D and somewhat saturates in comparison to the dynamics in 2BA. In Fig.~\ref{Fig_2D0D_0K}(b) we plot the occupation $N_{xx}$ after $0.8$~ps with both methods showing that a substantial negativity around $\epsilon_x=5~$meV builds up in 2BA (green triangles) while in 2BA-D (red dots) these strong negativities do not occur even though the density matrix is still not positive definite. This explains why the captured density increases without including the damping at longer times. Since the equations of motion in any case conserve the number of excitons, negativities in the occupation lead to an overestimation of other occupations which are usually the localized excitons, thereby overestimating their population in 2BA. This can also be seen in the spatial exciton density in Fig.~\ref{Fig_2D0D_0K}(c) where the incoherent density at $t=0.8~$ps is plotted in analogy to the case of Fig.~\ref{Fig_2D0D_0K}(b) where one can see an increased density within the potential at $R\approx 2~$nm and a small but spatially broad negativity for $5$~nm$\leq R\leq 15$~nm in the case of 2BA compared to 2BA-D. Also a very small negativity is build up for 2BA-D, but considerably smaller. This general feature is found also for the other studied potential and excitation conditions.

\section{Numerical details}
\label{App_numeric}
We here discuss the implementation of the energy-conserving $\delta$-function for scattering within delocalized states. For optical phonons no integral over a continuous variable affects the $\delta$-function such that we approximate the $\delta$-function by a Dirac sequence
	$\delta (\epsilon_x-E)\approx \delta_{\bar{\epsilon}_x}(\epsilon_x-E)$
with $\delta_{\bar{\epsilon}_x}(\epsilon_x-E)=\frac{1}{\sqrt{2\pi\bar{\epsilon}_x^2}}\exp\left(-\frac{(\epsilon_x-E)^2}{2\bar{\epsilon}_x^2}\right)$. We choose the broadening as
$$
\bar{\epsilon}_x=\mathrm{min}\left(\frac{|\epsilon_{x\pm 1}-\epsilon_x|}{4},\mathrm{1~meV}\right)
$$
such that in case of a quasi-resonant transition it is guaranteed that no states right above or below the resonant state strongly contribute. The cut-off energy of $1$~meV guarantees that not too large broadenings are considered. While in homogeneous systems the numerical energy separations are given by the momentum discretization $\Delta K$, it is here given by the simulated disk radius.  We checked that the above definition does not change the exciton dynamics considerably without confinement potential when considering the cases with a $R_{\mathrm{max}}=50$~nm and a $R_{\mathrm{max}}=100$~nm disk. Also the reduction of the cutoff from $1~$meV to $0.5~$meV did not change the results.\\
For scattering with acoustic phonons the $Q$-integration affects the $\delta$-function and can therefore be eliminated such that
	\begin{align*}
	&\int Q|g_{Q}|^2\left(K_{x'Q\bar{x}}\right)^*K_{\bar{x}Qx}\left(1+n_{Q}\right)\delta(\epsilon_{x}-\epsilon_{\bar{x}}+\hbar v_sQ)dQ\\
	&=\frac{\epsilon_{\bar{x}}-\epsilon_{x}}{\hbar^2v_{s}^2}|g_{Q_0}|^2\left(K_{x'Q_0\bar{x}}\right)^*K_{\bar{x}Q_0x}\left(1+n_{Q_0}\right)\Theta\left(\epsilon_{\bar{x}}-\epsilon_{x}\right)
	\end{align*}
with $Q_0=(\epsilon_{\bar{x}}-\epsilon_{x})/(\hbar v_{s})$ and the Heaviside function $\Theta$. Because of finite discretization $\Delta Q$, the value of the integrand at $Q_0$ is linearly interpolated.\\
For all given simulations we used a grid of 100 $Q$-points with a cutoff $Q_{\mathrm{max}}=6~\mathrm{nm^{-1}}$, $l_{\mathrm{max}}=80$ angular momentum states and a cutoff of $\epsilon_x\le 100~$meV for the simulations on a $R_{\mathrm{max}}=50~$nm disk.

\end{document}